\documentclass[epj]{svjour}

\usepackage{graphics}
\usepackage{dcolumn}
\usepackage{bm}
\usepackage{amsmath} 
\usepackage{hyperref}
\usepackage{amsfonts}
\usepackage{wasysym}
\usepackage{epsfig}    
\begin{document}
\title{Anomalous finite-size effects in the Battle of the Sexes}
\author{Jonas Cremer \and Tobias Reichenbach \and Erwin Frey
}                     
%
%
\institute{
Arnold Sommerfeld Center for Theoretical Physics (ASC) and
  Center for NanoScience (CeNS), Department of Physics,
  Ludwig-Maximilians-Universit\"at M\"unchen, Theresienstrasse 37,
  D-80333 M\"unchen, Germany} 

\date{Received: date / Revised version: date}
%
\abstract{The Battle of the Sexes describes asymmetric conflicts in mating behavior of males and females. Males can be philanderer or faithful, while females are either fast or coy, leading to a cyclic dynamics. The adjusted replicator equation predicts stable coexistence of all four strategies. In this situation, we consider the effects of fluctuations stemming from a finite population size. We show that they unavoidably lead to extinction of two strategies in the population. However, the typical time until extinction occurs strongly prolongs with increasing system size. In the meantime, a quasi-stationary probability distribution forms that is anomalously flat in the vicinity of the coexistence state. This behavior originates in a vanishing linear deterministic drift near the fixed point. We provide numerical data as well as an analytical approach to the mean extinction time and the quasi-stationary probability distribution.
} 
\maketitle

\section{Introduction}
Evolutionary game theory~\cite{Maynard,Hofbauer,Nowak} describes 
the coevolution of different interacting species.  The latter act according to a certain strategy, and the success of each strategy depends on the behavior of the other species. In this way, the fitness of each individual is frequency dependent; it changes upon altering the  abundances of the different species in the total population. The classical formulation of such evolutionary games uses  ordinary differential equations (replicator equations)~\cite{Maynard,Hofbauer}. However, in real populations fluctuations occur, stemming, e.g., from the discreteness of individuals, the random character of the interactions or spatial degrees of freedom. Incorporating such stochastic effects  constitutes an important issue that currently receives much attention~\cite{taylor,claussen-2005-71,traulsen-2005-95,traulsen-2006-74,reichenbach-2006-74,pacheco:258103,szabo-2007-446}. In the simplest scenario, fluctuations stem only from finite population sizes, while spatial structure or other correlations among individuals may be neglected. This situation arises when populations are well-mixed, i.e. individuals interact randomly with each other.

Here, we consider the influence of finite-size fluctuations on the dynamics of the bimatrix game \textit{Battle of the Sexes}. The latter was first introduced by Dawkin~\cite{Dawkins}, and serves as a model for describing  different mating behaviors of  males and females. Both males and females spread their genes if they get as many offspring as possible. However, an offspring causes some cost, and so every individual tries to shift this cost to the other mating partner, resulting in non-trivial competitions. As an example, this dynamics has been applied to insect populations~\cite{Parker}.  More generally, the Battle of the Sexes serves as an example of a intriguing class of asymmetric games, often also described by the game ``matching pennies'' ~\cite{Hofbauer,schustersigmund}. This class is characterized by lack of pure Nash-equilibria, accompanied by cyclic behavior and interesting stability properties of an emerging internal fixed point. 
  
In the simplest formulation of the Battle of the Sexes, both  males (\male) and females (\female) can choose between two alternative strategies.  Females either belong to the fast females, strategy $\female_A$, or to the coy females, strategy $\female_B$. The latter insist on a long courtship (implying a certain cost for both partners, as they wait for mating), opposed to fast females, while both care for their offspring (which also bears some cost). The male subpopulation is constituted of  philanderers, strategy $\male_A$,  or faithful males, strategy $\male_B$. Philanderers are not prepared to engage in a long courtship, and do not care for their offspring. In contrast, faithful males do, if necessary, engage in courtship as well as in raising the offspring.  Note that different strategies apply to the two subpopulations, the game is asymmetric.  The benefits and costs of the different strategies depend on the mating partner, and are usually encoded in a payoff matrix. For the Battle of the Sexes, Dawkin~\cite{Dawkins} has proposed the following payoffs:
\begin{displaymath}
 \label{eq:payoff}
 \begin{array}{l|ll}
  & \text{fast: }\female_A  & \text{coy: }\female_B  \\ \hline
 \text{philanderer: }\male_A& (15,-5) & (0,5) \\
 \text{faithful: }\male_B & (5,0)  & (2,2)
 \end{array}
 \end{displaymath}
Hereby,  the first term within the brackets encodes the payoff of males, and the second the one of females. 
For example, if a philanderer meets a fast female, the philanderer gains the high payoff $15$ (he has offspring without caring for it), while the fast female gets payoff $-5$ (although having spread her genes, she must raise the offspring herself). 

The qualitative behavior of the dynamics can be seen as follows.
Consider a situation with almost only fast females and philanderers, the few coy females will then take over the female subpopulation since fast females have a disadvantage against philanderers (they must raise the offspring themselves as the males leave after mating). However, if there are many coy females, faithful males are better off since fast females are rare and philanderers have problems finding mating partners. If there are again many faithful males, fast females have an advantage compared to coy ones since the risk of getting a philanderer is low and they safe the cost of waiting for mating. Philanderers, facing a high fraction of fast females, are now again in a favorable position, and edge out faithful males.   Hence, the whole dynamics is cyclic, similar to the non-hierarchical competition encountered in the rock-paper-scissors game~\cite{Maynard,Hofbauer,reichenbach-2006-74,szabo-2007-446,reichenbach-2007}.

The ecologically most relevant question in this competition is  whether all strategies can coexist (probably with oscillating frequencies) in a stable manner, such that the system is diverse, or whether some strategies go extinct, and only one pair of strategies survives. We  show in the following that the answer is drastically influenced by the finite sizes of the populations. If populations are large, the dynamics is close to the deterministic one. In this limit, it has been shown in~\cite{Maynard,schustersigmund} that all strategies coexist. The replicator equation predicts neutrally stable oscillations, and the adjusted replicator equation weakly damped ones. However, it has already been outlined in~\cite{traulsen-2005-95} that finite-size fluctuations may invalidate this result.

Here, we present  a thorough analysis of such stochastic effects in the Battle of the Sexes. Within a proper description of the system in terms of a Fokker-Planck equation, we show that extinction indeed occurs eventually, although the mean waiting time can be extremely long. Indeed, the latter increases strongly with growing population size. An intermediate probability distribution corresponding to coexistence of all strategies forms, which we compute analytically as well as by stochastic simulations. We find anomalous behavior; the distribution is flat over a wide range of states (corresponding to coexistence), and then strongly decays.

This paper is organized as follows: In the next section, we introduce the stochastic version of the Battle of the Sexes and present the master equation of the discrete stochastic process. 
In Section~\ref{sec:deterministic} we consider the deterministic limit of the stochastic model, and recover  adjusted replicator equations. We analyze the deterministic dynamics and  the stability of an interior coexistence fixed point, which gives us hints on the behavior of the stochastic model discussed in Sec.~\ref{sec:stochastic}. There, finite-size fluctuations are included into the description. We present the continuum approximation of the stochastic process by a Fokker-Planck equation and show that a quasi-stationary distribution forms. In Section~\ref{sec:extinction} we consider the mean extinction time $T$, being the mean time until two strategies go extinct. In the last Section we summarize our findings and present a brief conclusion.

\section{Stochastic model}
\label{sec:model}

In this Section we setup a simple stochastic model for describing the cyclic dynamics of the Battle of the Sexes in a population of finite  size. As we consider a well-mixed situation, the system's state is entirely determined by the number of individuals using the four different strategies. We denote by $N^{\male}_A$ and  $N^{\male}_B$ the number of males using strategy $\male_A$ resp. $\male_B$; analogously,  $N^{\female}_A$ and  $N^{\female}_B$ encode the number of females of strategy $\male_A$ resp. $\male_B$. 

For the  stochastic dynamics, we consider a Moran process~\cite{Moran,nowak-2005-433} (equivalent to urn models~\cite{reichenbach-2006-74,Feller}). In this framework, the number of males and females are chosen to be equal and remain constant in time, $N^{\female}=N^{\male}\equiv N$. Then we may, for example, describe the current state of the system by the frequencies $x$ of philanderers, $x=N^{\male}_A/N$, and the abundance $y$ of fast females, $y=N^{\female}_A/N$. The frequencies of faithful males and coy females follow  as $1-x$ and $1-y$. We thus encounter a two-dimensional  state space. The dynamics is defined by selection and reproduction events: At each update-step,  one individual is selected for reproduction, proportional to its fitness (which will be introduced below), while another (randomly chosen) individual of the same subpopulation is replaced.

Introducing the  probability $P(\vec x,t)$ that the system is in state $\vec x=(x,y)$ at time $t$, we can write down a master equation for the stochastic (Markov) process~\cite{Gardiner,Kampen},
\begin{equation}
\label{eq:master}
 \partial_t P\left(\vec x,t\right)=\sum_{\vec x'}\left[\Gamma_{\vec x'\to\vec x}P(\vec x',t)-\Gamma_{\vec x\to\vec x'}P(\vec x,t)\right]\,.
\end{equation}
Hereby, $\Gamma_{\vec x'\to\vec x}$ denotes the transition rate from state $\vec x$ to a new state $\vec x'$. In our case, the states can only differ by $1/N$, as only a single individual is replaced at each step. For example, a philanderer $(\male_A)$ can be replaced by a faithful male $(\male_B)$; we denote the corresponding rate by $\Gamma^{\male}_{A \to B}\left(\vec x\right)$ if the system's state was initially $\vec x$. Analogously, rates $\Gamma^{\male}_{B \to A}\left(\vec x\right)$, $\Gamma^{\female}_{A \to B}\left(\vec x\right)$  and $\Gamma^{\female}_{B \to A}\left(\vec x\right)$ appear. These transition rates involve the individuals' fitness, which we introduce in the following.

Denote by $f^{\male}_A$ the fitness of philanderers, and by $f^{\male}_B$, $f^{\female}_A$, $f^{\female}_B$ the ones of faithful males, fast females, and coy ones, respectively. They follow from a payoff matrix as the one presented in the introduction (to be described below). For the rates, as an example, consider the probability $\Gamma^{\male}_{A \to B}\left(\vec x\right)$ for replacing a philanderer by a faithful male. It is given by the fitness $f^{\male}_B$ of the latter (normalized by the average fitness denoted as $\phi^{\male}$), multiplied by the probability $x(1-x)$ of selecting a faithful male for reproduction and a philanderer for replacement. Thus, we encounter $\Gamma^{\male}_{A \to B}\left(\vec x\right)=\frac{f^{\male}_B}{\phi^{\male}}x(1-x)$.
Analogously, the other transition rates are found, and altogether, we obtain
\begin{eqnarray}
 \Gamma^{\male}_{B\to A}&=\frac{f^{\male}_A}{\phi^{\male}}x(1-x)\,,\quad
 \Gamma^{\male}_{A\to B}&=\frac{f^{\male}_B}{\phi^{\male}}x(1-x)\,,\cr
 \Gamma^{\female}_{B\to A}&=\frac{f^{\female}_A}{\phi^{\female}}y(1-y)\,,\quad\Gamma^{\female}_{A\to B}&=\frac{f^{\female}_B}{\phi^{\female}}y(1-y)\,.
\end{eqnarray}
Hereby, $\phi^{\female}$ denotes the average fitness of females, and is given by  $\phi^{\female}=f^{\female}_A y+f^{\female}_B (1-y)$, while the average fitness of males  reads $\phi^{\male}=f^{\male}_A x+f^{\male}_B (1-x)$.
The rate for a transition within a given  subpopulation depends on the frequencies of strategies there, as well as  on the abundances of the different strategies of the other subpopulation, entering via the fitness functions, as we show in the following.

As described in the introduction, fitness is conveniently encoded in a payoff matrix. Here, we consider the most symmetric formulation of  the Battle of the Sexes, where the payoffs are given by
\begin{displaymath}
 \label{eq:payoff}
 \begin{array}{l|ll}
  & \text{fast: }\female_A  & \text{coy: }\female_B  \\ \hline
 \text{philanderer: }\male_A& (1,-1) & (-1,1) \\
 \text{faithful: }\male_B & (-1,1)  & (1,-1)
 \end{array}
 \end{displaymath}
For example, if a philanderer meets a fast female, the philanderer gains payoff $1$, while the fast female gets payoff $-1$. 

For the fitness of one individual, we consider a background fitness of $1-\omega$, where the (small) parameter $\omega$ is referred to as selection strength. The fitness stemming from  the individuals' interactions, given by the payoff matrix, is multiplied by $\omega$, and added. For vanishing selection strength, $\omega=0$, the game is purely neutral, there is no frequency dependent fitness at all. Increasing $\omega$, the interactions between the individuals and therefore the frequency dependent dynamics becomes more and more important. For specificity, in our stochastic simulations presented below, we used $\omega =0.5$. 
 
In this formulation, the fitness of a  philanderer is $f^{\male}_A=1-\omega+\omega\left[y-(1-y)\right]$, as he encounters a fast female at probability $y$ and a coy one with a probability $1-y$. Similarly, faithful males gain a payoff  $f^{\male}_B=1-\omega+\omega\left(1-2y\right)$, fast females earn $f^{\female}_A=1-\omega+\omega\left(1-2x\right)$, and coy ones $f^{\female}_B=1-\omega+\omega\left(2x-1\right)$.

\section{Deterministic dynamics}
\label{sec:deterministic}

To gain a first intuitive understanding of the system's behavior, we start our discussion with the deterministic dynamics, which is ensued in the limit of infinite population size, $N\to\infty$. In this asymptotic limit, the state space becomes continuous: the frequencies of philanderers, $x$, and  of fast females, $y$, can take any value in the interval $[0,1]$. 
The deterministic dynamics describes the time-evolution of these frequencies  in terms of ordinary differential equations (ODE). The latter are obtained from the underlying stochastic process by consideration of the frequencies' mean values. In addition, a mean-field assumption (which is justified for well-mixed populations) factorizes higher moments of the probability distribution, yielding ODE for the first moments.  For the Moran process, which we consider, they are given by the adjusted replicator equations~\cite{traulsen-2005-95} for the (mean) frequencies $x$ and $y$, we sketch its derivation in the following.    According to the Moran process introduced above, the time evolution of the mean fraction of philanderers is given by
\begin{equation}
 x(t+\delta t)=x(t)+\frac{f^{\male}_A}{\phi^{\male}}x(1-x)\delta t-\frac{f^{\male}_B}{\phi^{\male}}x(1-x)\delta t.
\end{equation}
Here, we have chosen the time-scale such that $N$ update steps (one generation) occur in one time-unit.
Together with an analogous equation for the time evolution of the females' frequencies, we obtain
\begin{eqnarray}
\label{eq:adjustedreplicator}
 \partial_t x&=&\frac{f_A^{\male}-\phi^{\male}}{\phi^{\male}}x=2\tilde\omega\frac{x(1-x)(2y-1)}{1+\tilde\omega\left(2y-1\right)\left(2x-1\right)}\,,\nonumber\\
 \partial_t y&=&\frac{f_A^{\female}-\phi^{\female}}{\phi^{\female}}y=-2\tilde\omega\frac{y(1-y)(2x-1)}{1-\tilde\omega\left(2x-1\right)\left(2y-1\right)}\,, 
\end{eqnarray}   
where we introduced $\tilde \omega=\frac{\omega}{1-\omega}$ to simplify the expressions.
We recover the well known adjusted replicator equations of a bimatrix game~\cite{Hofbauer,hofbauerbimatrix}. The success of a strategy depends on its fitness  compared to the average one. For example, the sign of $f_A^{\male}-\phi^{\male}$ determines if the fraction of philanderers in the male subpopulation grows or decreases.

Let us  analyze the deterministic system described by Eqs.~(\ref{eq:adjustedreplicator}) in more detail.
There are four trivial fixed points, namely the four corners of the phase space. They correspond to the survival of only one pair of strategies, and are given by $(x^*_1,y^*_1)=(0,0)$ (only faithful males and coy females), $(x^*_2,y^*_2)=(0,1)$ (only faithful males and fast females), $(x^*_3,y^*_3)=(1,0)$(only philanderers and coy females), and $(x^*_4,y^*_4)=(1,1)$  (only philanderers and fast females). All these fixed points are saddle points. In addition,  one fixed point resides in the interior of the phase space, at a position  $x^*=y^*=1/2$. At this coexistence fixed point,  all strategies are present at equal frequencies.

The coexistence fixed point is globally stable, which can be shown using the following Lyapunov function:
\begin{eqnarray}
\label{eq:lyapunovfunction}
 H&=&-x\cdot(1-x)\cdot y\cdot(1-y)\\ \nonumber
&=&-\left(\frac{1}{4}-\tilde x^2\right)\left(\frac{1}{4}-\tilde y^2\right)\,.
\end{eqnarray}
Hereby, in the second line, we have  introduced new coordinates $\tilde x,\tilde y$ that originate in the coexistence fixed point, 
\begin{eqnarray}
\tilde x=x-{1\over2}\nonumber\,, \quad \tilde y=y-{1\over2}.
\end{eqnarray} 
In these coordinates, the  temporal derivative of $H$ is given by 
\begin{equation}
\partial_t H=-\frac{\tilde \omega^2\tilde  x^2\tilde  y^2 4\left(1-4\tilde  x^2 \right) \left( 1-4\tilde  y^2\right)}{\left(1-\tilde \omega\right)^2-\left( 4\tilde\omega\tilde   x \tilde  y \right)^2},
\end{equation}
which is negative on the whole phase space (except for the boundaries and the coexistence fixed point, where it vanishes). As $H$ has its minimum at the coexistence fixed point, the latter is globally stable. In particular, it is attractive for every state that is not located at the boundaries.

Performing a Taylor expansion  of the Eqs.~(\ref{eq:adjustedreplicator}) around the fixed point yields
\begin{eqnarray}
 \label{eq:approximatedreplicator}
  \partial_t \tilde x&=&\tilde \omega\tilde y -4\tilde\omega \tilde x^2\tilde
 y -4\tilde \omega^2\tilde x\tilde y^2+\mathcal{O}\left(\tilde x^2,\tilde y^2
 \right)\,,\nonumber\\
  \partial_t \tilde y&=&-\tilde \omega\tilde x -4\tilde\omega^2 \tilde
 x^2\tilde y +4\tilde \omega\tilde x\tilde y^2+\mathcal{O}\left(\tilde
 x^2,\tilde y^2 \right)\,.
 \end{eqnarray}
 The dynamics resulting from these equations becomes most evident 
upon introducing polar coordinates $(r,\phi)$ given by 
\begin{eqnarray}
\tilde x=r\cos \phi \text{, and }\tilde y=r\sin \phi. 
\end{eqnarray}
$r$ denotes the distance to the coexistence fixed point. Then, the replicator dynamics is given by
\begin{eqnarray}
\label{eq:polarreplicator}
  \partial_t r&=&-\tilde \omega
 r^3\left[\tilde \omega(1-\cos4\phi)+\sin4\phi
 \right]\nonumber,\\
  \partial_t \phi&=&-\tilde \omega +\tilde \omega
 r^2\left[(1-\cos 4\phi)-\tilde \omega \sin 4\phi\right]) .
 \end{eqnarray}
We observe  the linear order to predict neutrally stable cycling around the fixed point at frequency $\tilde \omega$. Therefore, this behavior occurs asymptotically when approaching the coexistence fixed point. At further distance, higher orders matter. As can be seen, the third order terms in $\partial_t r$ yield a drift towards the fixed point\footnote{The average of the angle-dependent terms in $\partial_t r$ along one cycle around the fixed point vanishes, the remaining term yields a drift into the coexistence fixed point.}. The latter is thus stable, in agreement with the previous considerations based on the Lyapunov function.

\begin{figure}
\begin{center} 
\includegraphics[width=8cm]{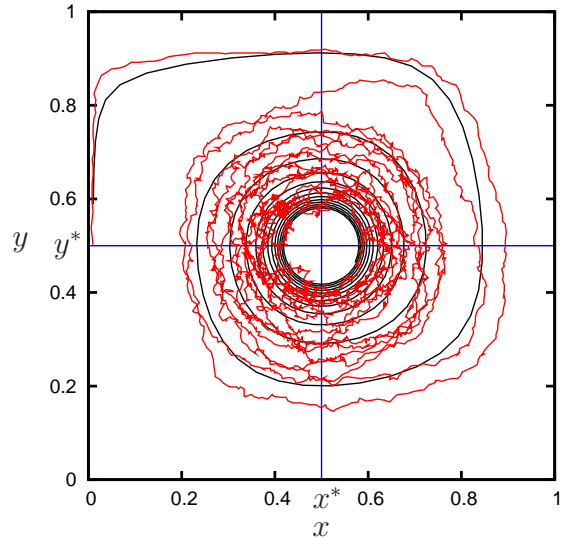} 
\end{center}
\caption{Phase space and dynamics of the Battle of the Sexes. The deterministic dynamics exhibits a coexistence fixed point at $x^*=y^*=\frac{1}{2}$. The black solid line shows a deterministic trajectory.  Starting at a fraction of philanders of $x=0.01$ and a fraction of fast females given by $y=0.5$, the system oscillates according to the cyclic dynamics. The coexistence fixed point is approached but the drift towards it is getting very weak in its vicinity. Indeed, the linear order of the drift vanishes at the fixed point, and only third order terms render it stable. A realization of the stochastic process (starting at the same initial state; the system size is $N=500$) is shown in red. As can be seen, at the beginning, the realization follows the deterministic trajectory. The drift towards the coexistence fixed point dominates the behavior, fluctuations are unimportant. This changes at closer distance to the coexistence fixed point, where the  deterministic drift is lower and fluctuations have  a stronger impact.}
\label{fg:btsdet}
\end{figure}
To summarize, although the coexistence fixed point is globally stable, the linear drift in its vicinity vanishes. Therefore, close to the fixed point, only cyclic motion around it remains. To illustrate this behavior, 
in Fig.~\ref{fg:btsdet}, we show  a trajectory (black solid line) emerging as solution of the adjusted replicator equation when starting near the boundary of the phase space. Although the drift towards the interior fixed point is strong close to the boundary, it is seen to vanish when the fixed point is approached. There, the cyclic behavior dominates. As an example for the behavior of the finite system (discussed in the next Section), we also show a stochastic trajectory (red line) in Fig.~\ref{fg:btsdet}. Starting in the vicinity of the boundary, it follows closely the deterministic solution. The deterministic drift is strong, and fluctuations have only a minor influence. This changes when the fixed point is approached. There, due to the vanishing deterministic drift towards the fixed point, fluctuations dominate.    In the next Section, we show how this induces intriguing  behavior.

\section{Stochastic description}
\label{sec:stochastic}

Stochasticity has an important impact in this model due to the existence of absorbing boundaries. Namely, if one strategy has gone extinct (i.e. the system has reached the boundary of the phase space), this strategy can no longer give rise to an offspring, and therefore cannot recover again. Thus, the boundaries are absorbing, and once the system has reached them, the cyclic dynamics drives it into a state where only one pair of strategies is present (the corners of the phase space).  The existence of fluctuations in the system ensures that (maybe after very long waiting time) the absorbing boundary is reached. Thus, extinction unavoidably occurs. However, due to the deterministic drift towards the coexistence fixed point, the mean waiting time $T$ is expected to be long for large systems. Indeed, in the next Section, we will show that $T$ strongly prolongs with increasing population size. 

\begin{figure}
\begin{center}
\includegraphics[width=8cm]{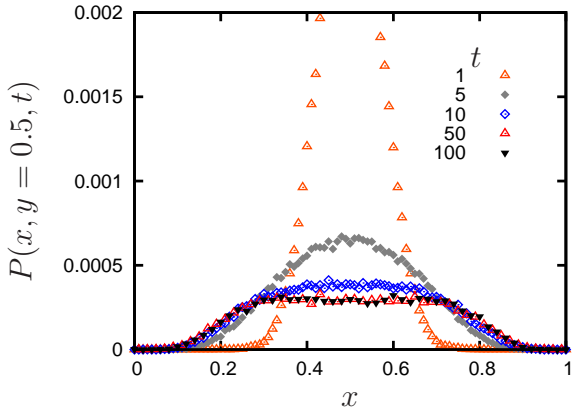}
\end{center}
\caption{The approach of a quasi-stationary distribution. Stochastic simulations (for a system size $N=100$) show that, starting at an initial  delta peak at the coexistence fixed point, a quasi-stationary distribution is reached, at a time around $t=50$. Then, it remains unchanged for a long time (we have followed it up to a time $t=100$).  During this time, extinction events have occurred only very rarely (until $t=100$, less then 0.1\% of the realizations have reached the boundaries). We observe the quasi-stationary distribution to be almost flat in the vicinity of the reactive fixed point. }
\label{fg:btsnvar1}
\end{figure}

\begin{figure}
\begin{center}
\includegraphics[width=8cm]{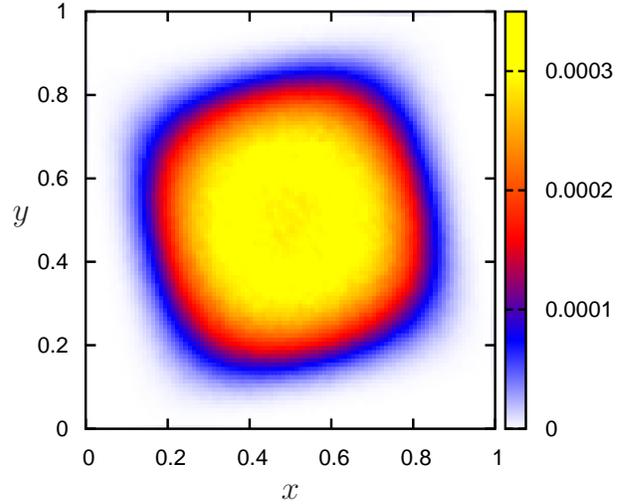}
\end{center}
\caption{The quasi-stationary distribution, for a system size $N=100$. It is concentrated in the middle of the phase space, having almost flat behavior there. Towards the boundaries, it rapidly decreases, until  nearly vanishing. 
The probability distribution is symmetric under rotation of $\pi/2$, as follows also from the symmetric formulation of the game (symmetric payoff matrix). Close to the fixed point, it asymptotically obeys a polar symmetry. For large system size, the relevant part of the quasi-stationary distribution is located close enough around the fixed point. In this regime, the approximation by an angular independent Fokker-Planck equation is valid. }
\label{fg:btsnvar2}
\end{figure}

Consider a situation where the system is initially at the coexistence fixed point (the probability distribution is, initially, a delta peak). The distribution will broaden in time, counteracted by deterministic third-order terms, c.f. Fig.~\ref{fg:btsnvar1}. When this deterministic drift balances the fluctuations, an intermediate (quasi-stationary) probability distribution forms (Fig.~\ref{fg:btsnvar2}). As the initial spreading is induced by fluctuations, the intermediate distribution is expected to form around a time proportional to the system size $N$~\cite{reichenbach-2006-74}. (In this section we show that it obeys non-gaussian behavior, due to the vanishing linear drift.) At much later times, this distribution decays as, due to fluctuations, the probability increases that the system has eventually reached the absorbing boundaries. We find the typical extinction time $T$ to diverge exponentially in the system size, $T\sim\exp(N)$. 

In this Section we describe the quasi-stationary distribution analytically; the mean extinction time is considered in the next Section.
To take the stochasticity of the Moran process into account, we chose a continuous diffusion-like description. Namely, we use a Fokker-Planck equation obtained via a Kramers-Moyal expansion (which may be seen as an expansion in the system size) of the master equation~(\ref{eq:master}), see , e.g., Ref.~\cite{taeuber}. The resulting Fokker-Planck equation describes the time evolution of the probability $P(\vec x,t)$ that the system is in state $\vec x=(x,y)$ at time $t$, and reads
\begin{eqnarray}
\label{eq:FPE}
 \partial_t P(\vec x,t)&=&-\sum_{i}\partial_{i}\left[ \alpha_i(\vec x,t)P(\vec x,t)\right]\nonumber\\
& &+{1\over 2}\sum_{i,j}\partial_{i}\partial_{j}\left[\beta_{ij}P(\vec x,t)\right].
\end{eqnarray}
The summation is over the phase space coordinates $x$  and $y$. 
The drift coefficients $\alpha_i$ follow from the adjusted replicator equations~(\ref{eq:adjustedreplicator}); $\alpha_x=\partial_t x$ and $\alpha_y=\partial_t y$. 
The coefficients of the diffusion matrix $(\beta_{ij})$ encode fluctuations and are given by
\begin{eqnarray}
\beta_{xx}&=&\frac{1}{N}\left(\frac{f_A^{\male}}{\phi^{\male}}x(1-x)+\frac{f_B^{\male}}{\phi^{\male}}(1-x)x\right),\nonumber\\
\beta_{yy}&=&\frac{1}{N}\left(\frac{f_A^{\female}}{\phi^{\female}}y(1-y)+\frac{f_B^{\female}}{\phi^{\female}}(1-y)y \right),\nonumber\\
\beta_{xy}&=&\beta_{yx}=0.
\end{eqnarray} 
The diffusion matrix $(\beta_{ij})$ is diagonal, as can be understood as follows: Each update process is only acting on one subpopulation, such that fluctuations between the subpopulations do not form. For each subpopulation, a process leading to an increase or decrease of one strategy increases the fluctuations, therefore, for example, $\beta_{xx}$ is the sum of the probability for an increase of the philanderers and for an increase of the faithful males. 

To proceed, we further employ van Kampens linear noise  approximation~\cite{Kampen}. We assume a constant diffusion matrix with the values at the coexistence fixed point, $\beta_{ij}(\vec x)\approx\beta_{ij}(\vec x^*)$. In addition, as done above in the deterministic description, we approximate the drift term up to third order around the coexistence fixed point and use polar coordinates. The Fokker Planck equation turns into
\begin{eqnarray}
\partial_t P(r,\phi,t)&=&\tilde \omega \partial_{\phi}P(r,\phi,t)+4\tilde \omega^2  r^2 P(r,\phi,t)\nonumber\\
& &+~\tilde \omega r^3 \left[ \tilde \omega\left( 1-\cos 4\phi \right)+\sin 4 \phi  \right]\partial_r P(r,\phi,t) \nonumber\\
& &+~\tilde \omega r^3 \left[  \tilde \omega \sin 4\phi-\left( 1-\cos 4\phi \right) \right]\partial_{\phi}P(r,\phi,t)\nonumber \\
& &+{1 \over 2N} \bigtriangleup_r P(r,\phi,t),
\end{eqnarray}
where $\bigtriangleup_r$ denotes the Laplacian in polar coordinates.\\
As can be seen from Fig.~\ref{fg:btsdet} as well as from Eqs.~(\ref{eq:polarreplicator}), in the vicinity of the coexistence fixed point, the system obeys a polar symmetry. Therefore, we solve the Fokker-Planck equation  within a radial approximation around the fixed point by replacing every angle-dependent term by its mean value, for example $\langle \sin \phi \rangle_{\phi}=0$. This is justified as close to the fixed point, the system performs many oscillations with only weak drift into the fixed point. The drift may therefore be approximated by its mean value on a given cycle. 
With $\tilde P(r)=r \cdot P(r,t)$ it follows:
\begin{eqnarray}
\label{eq:btsapproxtilde}
\partial_t \tilde P(r,t)&=&\partial_r \left[\left( \tilde\omega^2 r^3 -{1\over 2N}{1 \over r}\right)\tilde P(r,t)\right]+{1\over 2N}\partial^2_{r}\tilde P(r,t)\,,\cr
&=&-\partial_r\left[\alpha_r\tilde{P}(r,t)\right]+\frac{1}{2}\beta_r\partial_r^2\tilde{P}(r,t)\,.
\end{eqnarray}
We encounter a one-dimensional Fokker-Planck equation. The dynamics takes place in an effective potential $U={1\over 4}\tilde \omega^2r^4-{1\over 2N}\text{ln}(r)$, with $\alpha_r=-\partial_r U(r)$. Standard diffusion, given by the constant fluctuation term $\beta_r={1\over N}$, occurs. With the help of this approximation we can now calculate the quasi stationary distribution defined by $\partial_t P(r,t)=0$. For large enough systems, the distribution is centered at the fixed point, see Fig.~\ref{fg:btsnvar2}, nearly vanishing at the boundaries. We therefore neglect the absorbing boundaries of the phase space, and use $P(\infty,t)=0$ as boundary condition. The radial quasi-stationary distribution then reads
\begin{equation}
P(r)=\frac{2d}{ \pi^\frac{3}{2}}\exp \left(-d r^4 \right),
\end{equation}
with $d=\frac{\tilde \omega^2 N}{2}$. Hereby, $r$ is the distance to the coexistence fixed point.

\begin{figure}
\begin{center}
\includegraphics[width=8cm]{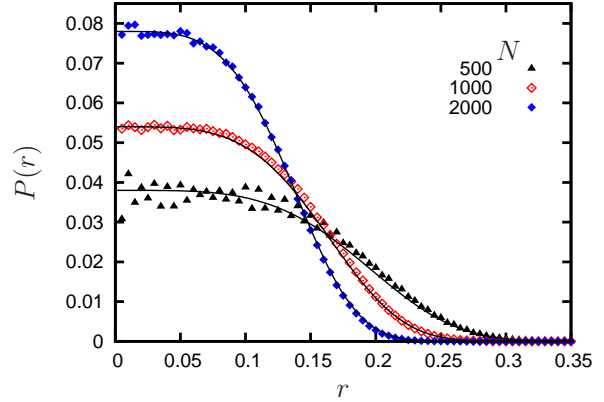}
\end{center}
\caption{Stochastic simulations of the  Battle of the Sexes. Numerical data for the radial quasi-stationary probability distribution are shown. They confirm the analytical prediction of an $\exp({-dr^4})$-distribution (solid lines). With increasing  system size $N$, the distribution is more and more centered around the fixed point. In the vicinity of the latter, the distribution is nearly flat, but it falls rapidly to zero at a larger distance.}
\label{fg:btsnvar3}
\end{figure}
We have compared this analytical finding to stochastic simulations, see Fig.~\ref{fg:btsnvar3}. Both agree excellently and thereby the approximation by an angle independent probability distribution is confirmed.

This anomalous probability distribution is a result of the weak deterministic drift towards the boundaries. Indeed, a linearly stable fixed point leads to a (standard) gaussian distribution, with a standard deviation that decreases as $N^{-1/2}$ in the system size $N$, confer to an Ornstein-Uhlenbeck-process~\cite{Gardiner,Kampen}. In our case, the linear drift at the fixed point vanishes, consequently, the resulting probability  distribution is almost flat there. At larger distance to the fixed point, third order terms become important, and cause a sudden decrease of the probability; a non-gaussian distribution forms. Its width, encoded by the parameter $d^{-1/4}$, still decreases with increasing system size $N$,  although weaker than in a gaussian distribution, namely only as $N^{-1/4}$. This behavior universally arises when a fixed point is neutrally stable to linear order, but stabilized by third order terms, and therefore characterizes this class of systems.

\section{Extinction time}
\label{sec:extinction}

In the previous section, we have described the emergence of an intermediate quasi-stationary probability distribution on time scales proportional to the system size. However, we have already pointed out that, on much longer time scales, this distribution eventually decays, and only one pair of strategies survives (in our model, due to the symmetric formulation, all corners of the phase space may be reached at equal probabilities). In this section, we analyze the time scale needed for extinction. In particular, we show that it diverges exponentially in the system size $N$. 
 
Consider  the mean time $T(x,y)$ it takes until the absorbing boundaries are reached if starting at a state $(x,y)$ at time $t=0$. For its calculation, we use again the continuous approximation of the stochastic process by the Fokker-Planck equation~(\ref{eq:FPE}) introduced above. Also,  we employ the angular symmetric approximation to obtain the one-dimensional Fokker-Planck equation~(\ref{eq:btsapproxtilde}). In this setting, calculating the extinction time is  a first-passage time problem in one dimension with an absorbing boundary at a radius $r_{\text{exit}}$. We are interested in the asymptotic limit of a large system size, and for the scaling behavior of $T$, the explicit value of $r_{\text{exit}}$ will turn out  be unimportant. 
We solve the first-passage problem with the help of the backward Kolmogorov equation~\cite{Gardiner,Risken}. Thereby, the mean extinction time is given by the solution of the following equation
\begin{equation}
\left[\alpha_r(r)\partial_r+{1\over 2}\beta_r(r)\partial_r^2\right]T(r)=-1,
\end{equation}
with appropriate boundary conditions. The variable $r$ now denotes the starting point, at time  $t=0$. The general solution of this problem for one reflecting ($r=0$) and one absorbing boundary ($r=r_{\text{exit}}$) is given by
\begin{equation}
T=2\int_{r}^{r_{\text{exit}}}{dy\over \Psi(y)}\int_0^y{\Psi(z)\over \beta_r} dz\,.
\label{eq:Tint}
\end{equation}
Hereby,  $\Psi(z)=\exp\left[\int_0^z dz' {2\alpha_r(z')\over \beta_r} \right]$, and is computed as
\begin{equation}
 \Psi(z)=z\exp\left(-\frac{N\tilde \omega^2}{2}z^4 \right)\,.
\end{equation} 
We consider the mean extinction time when starting at the coexistence fixed point, $T=T(r=0)$.
Then, the integrals in Eq.~(\ref{eq:Tint}) are solved by a hypergeometric series:
\begin{equation}
\label{eq:exttime}
T={N\sqrt{\pi}\over 2}r_{\text{exit}}^2\text{F}_{\left\lbrace 1/2,1;3/2,3/2 \right\rbrace }\left(\frac{N\tilde\omega^2}{2} r_{\text{exit}}^4\right)\,.
\end{equation}
Hereby, the  hypergeometric series is defined as
\begin{eqnarray}
\text{F}_{\left\lbrace a_1,a_2;b_1,b_2\right\rbrace }\left( z\right)=\sum_{k=0}^\infty \frac{\Gamma(a_1+k) \Gamma(a_2+k)}{\Gamma(a_1) \Gamma(a_2)}\times\nonumber\\ \frac{\Gamma(b_1) \Gamma(b_2)}{\Gamma(b_1+k) \Gamma(b_2+k)}{z^k\over k!}\,.
\end{eqnarray}
Equation~(\ref{eq:exttime}) yields  an analytical expression for the mean extinction time. In the asymptotic limit of  infinite system size $N$, it turns out that $T$ diverges exponentially, $T\sim N^{-1/2} \exp(N)$. Namely, an asymptotic expansion of the hypergeometric series for large arguments yields $\text{F}_{\left\lbrace a_1,a_2;b_1,b_2\right\rbrace }\left( z\right)\sim z^{a_1+a_2-b_1-b_2}\exp(z)$, see, e.g.,~\cite{functions.wolfram}. For the dependence of the mean extinction time $T$ on $N$, this implies $T\sim N\times N^{a_1+a_2-b_1-b_2}\exp(N)=N^{-1/2}\exp(N)$.

\begin{figure}
\begin{center}
\includegraphics[width=8cm]{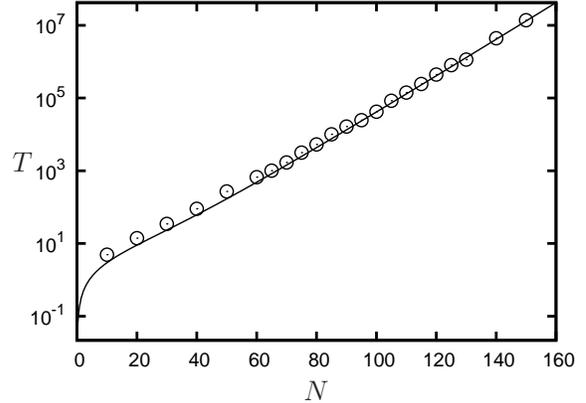}
\end{center}
\caption{The mean extinction time $T$: Stochastic simulations (circles) show that $T$  increases exponentially with increasing system size $N$. This originates in finite-size fluctuations that, with larger system size,  become weaker and weaker compared to the deterministic drift towards the coexistence fixed point. The analytical approximation, given by Eq.~(\ref{eq:exttime}), is shown as a solid line and agrees very well with the numerical data. We have chosen an exit radius of $r_\text{exit}=0.70$, and adjusted an additional factor of $0.28$, taking nonlinearities into account.}
\label{fg:btsextinction}
\end{figure}
In Fig.~\ref{fg:btsextinction} the analytical prediction of the extinction time as well as stochastic simulations are shown. Hereby, excellent agreement is found for  $r_\text{exit}=0.70$. This value lies in between the minimal possible exit radius, $r_\text{exit}^\text{min}=0.5$, and the maximal possible one (the distance from the coexistence fixed point to one corner), $r_\text{exit}^\text{max}=1/\sqrt{2}$. In addition, we have adjusted a constant factor of $0.28$, taking nonlinearities into account\footnote{By assuming a constant fluctuation term and approximating the deterministic drift term up to third order we have neglected higher nonlinearities. These higher order terms do not change the characteristic behavior of the extinction time but lead to an additional factor (in our case $0.28$), a similar example is considered in~\cite{Jonas}.}.

\section{Conclusions}

In this article, we have considered the influence of finite-size fluctuations on the coexistence of different strategies in  the Battle of the Sexes. The deterministic dynamics, described by adjusted replicator equations, predicts a coexistence fixed point which is stable (in contrast to the standard replicator dynamics considered in~\cite{schustersigmund}, which yields neutrally stable oscillations).  However, the linear part of the deterministic drift towards the coexistence fixed point vanishes, stability is only induced by third order terms. We have shown how this behavior, in the presence of finite-size fluctuations that compete with the deterministic drift, induces a non-gaussian probability distribution, proportional to $\exp(-r^4)$ around the fixed point. Namely, as the drift vanishes in the vicinity of the coexistence fixed point, fluctuations effect the system unopposed and induce a flat distribution there. Further away from the fixed point, third order terms of the deterministic drift dominate and lead to a sudden decrease of the probability distribution.

We have found this probability distribution to be only intermediate. Namely, due to the presence of absorbing boundaries, extinction of two strategies eventually occurs. We have computed the mean time for extinction, and found that it increases exponentially in the  system size $N$. Thus, for large systems, the mean extinction time is extremely long, and the quasi-stationary probability distribution very long-lived. This behavior originates in the $1/N$ decay of the finite-size fluctuations, rendering the deterministic drift dominant for large system sizes. On time-scales below the mean extinction time, which may be the ecologically relevant ones (see, e.g., the discussion in~\cite{hastings-2001-4,hastings-2004-19}), coexistence is thus stable. Indeed, in Ref.~\cite{reichenbach-2007} a definition of stability/instability of coexistence in stochastic systems based on time-scales has been proposed; within this classification, the stochastic Battle of the Sexes analyzed in the present work belongs to the stable regime. We hope that, in such a more general context,  our work  contributes to the understanding of ecological diversity in the  interplay with fluctuations and time-scales.
 
Although we have, for specificity, considered a symmetric formulation of the Battle of the Sexes, on a qualitative level, our results generalize to less symmetric cases. Indeed, as long as the payoff matrix belongs to the class of the game ``matching pennies'', its precise entries have only minor influence. The deterministic (adjusted replicator) equations obey a cyclic dynamics, with an internal fixed point that is neutrally stable to first order, while higher orders yield a drift towards it. In this situation, finite-size fluctuations  cause the behavior described above - extinction on an exponentially diverging time-scale, with an intermediate anomalous probability distribution. Considering situations where individuals are spatially arranged~\cite{szabo-2007-446,reichenbach-2007,nowak-1992-359} or interact with each other according to network structures (see e.g.~\cite{pacheco:258103,szabo-2007-446,colizza-2007}), one may speculate that the intriguing dynamics of this class of games may lead to fruitful further insight into the role of fluctuations and correlations on ecosystems diversity.

\vspace{0.5cm}
Financial support of the German Excellence Initiative via the program ``Nanosystems Initiative Munich (NIM)'' is gratefully acknowledged.

%
%
%



\end{document}